\documentclass[conference]{IEEEtran}
\IEEEoverridecommandlockouts
\usepackage{cite}
\usepackage{amsmath,amssymb,amsfonts}
\usepackage{graphicx}
\usepackage{textcomp}
\usepackage{xcolor}

\usepackage{amsfonts,amssymb}
\usepackage{amsmath, bm}
\usepackage{mathrsfs}
\usepackage{algorithm}
\usepackage{algorithmicx}
\usepackage{algpseudocode}
\usepackage{makecell}
\usepackage{subfigure}

\usepackage{braket}

\def\BibTeX{{\rm B\kern-.05em{\sc i\kern-.025em b}\kern-.08em
    T\kern-.1667em\lower.7ex\hbox{E}\kern-.125emX}}
\begin{document}
	
\newcommand{\B}{\mathbf}
\newcommand{\BS}{\boldsymbol}
\newcommand{\T}{^\mathsf{T}}
\newcommand{\CT}{^\mathsf{H}}

\title{Successive Pose Estimation and Beam Tracking for mmWave Vehicular Communication Systems}

\author{
	\IEEEauthorblockN{Cen Liu$^*$, Guangxu Zhu$^\dag$, Fan Liu$^\ddag$, Yuanwei Liu$^\mathsection$ and Kaibin Huang$^\mathparagraph$}
	\IEEEauthorblockA{$^*$National University of Singapore, Singapore\ \ $^\dag$Shenzhen Research Institute of Big Data, Shenzhen, China\\
	$^\ddag$Southern University of Science and Technology, Shenzhen, China\ \ $^\mathsection$Queen Mary University of London, London, U.K.\\
	$^\mathparagraph$The University of Hong Kong, Hong Kong\\
	Email: liucen@u.nus.edu, gxzhu@sribd.cn, liuf6@sustech.edu.cn, yuanwei.liu@qmul.ac.uk, huangkb@eee.hku.hk}
}

\maketitle

\begin{abstract}
The millimeter wave (mmWave) radar sensing-aided communications in vehicular mobile communication systems is investigated. To alleviate the beam training overhead under high mobility scenarios, a successive pose estimation and beam tracking (SPEBT) scheme is proposed to facilitate mmWave communications with the assistance of mmWave radar sensing. The proposed SPEBT scheme first resorts to a Fast Conservative Filtering for Efficient and Accurate Radar odometry (Fast-CFEAR) approach to estimate the vehicle pose consisting of 2-dimensional position and yaw from radar point clouds collected by mmWave radar sensor. Then, the pose estimation information is fed into an extend Kalman filter to perform beam tracking for the line-of-sight channel. Owing to the intrinsic robustness of mmWave radar sensing, the proposed SPEBT scheme is capable of operating reliably under extreme weather/illumination conditions and large-scale global navigation satellite systems (GNSS)-denied environments. The practical deployment of the SPEBT scheme is verified through rigorous testing on a real-world sensing dataset. Simulation results demonstrate that the proposed SPEBT scheme is capable of providing precise pose estimation information and accurate beam tracking output, while reducing the proportion of beam training overhead to less than $5\%$ averagely.
\end{abstract}


\section{Introduction}
In the past few years, millimeter wave (mmWave) technology and its promising applications have been getting growing interest from both wireless communications and robotics communities. On one hand, mmWave communications is known as a key technology in 5G New Radio (NR) and beyond 5G cellular network \cite{Rangan14, LiliWei14}. The extremely high frequency (EHF) band of mmWave is able to provide enormous amount of spectrum resources to meet the needs of diverse user applications. On the other hand, mmWave radar sensing plays an indispensable role in modern autonomous robotic systems. Owing to its intrinsic robustness towards bad illumination and harsh weather conditions, mmWave radar sensing is capable of supplying more stable and reliable perception information than vision and LiDAR solutions under global navigation satellite systems (GNSS)-denied environments \cite{SarahCen18, ZiyangHong20, YinZhiNg21}.

To counteract high attenuation of free space propagation as well as blockage penetration in EHF band, mmWave communications generally resorts to large-scale multi-input multi-output (MIMO) antennas in forming high directional beams aligned to the wireless terminals via transmit precoding and receive beamforming \cite{Alkhateeb14}. Nevertheless, it tends to incur unbearable beam training overhead in vehicular mobile communication systems since the vehicle moves at a high speed and thus the position of mobile station (MS) changes intensively. To address this issue, beam tracking serves as a commonly-adopted technique to alleviate the burden incurred by frequent beam training \cite{YiWang22}.

Several works related to the beam tracking in mmWave mobile communication systems have emerged in recent years \cite{ChuangZhang16, Va16, Jayaprakasam17, FuliangLiu19, Shaham20, LiChen23}. The authors of \cite{ChuangZhang16} proposed a dual-timescale model on characterizing abrupt channel changes as well as tracking slow variations of angles of departure (AoDs) and angles of arrival (AoAs) for multiple dominant paths by assuming the path gains are priori known. On the contrary, in \cite{Va16}, the AoD/AoA and the path gain of a single line-of-sight (LoS) path are tracked using extended Kalman filter (EKF) algorithm through training a single beam pair at each timeslot. Assuming parallelism between the antenna arrays of base station (BS) and MS, a joint minimum mean square error (MMSE) robust beamforming and EKF tracking scheme is proposed in \cite{Jayaprakasam17} to minimize the mismatch on beamforming angles between BS and MS. In \cite{FuliangLiu19}, the authors proposed a second-order AoD/AoA variation model by taking the slow variations of AoD/AoA angular velocity into account. \cite{Shaham20} investigated the EKF-based beam tracking approach for vehicular communication systems by incorporating 1-dimensional (1D) position and velocity of the vehicle into state variables instead of tracking AoD/AoA directly. The recent work \cite{LiChen23} utilized 2-dimensional (2D) position and angular velocity of MS to back up the EKF tracking approach under high mobility scenarios, and offered insights towards the formulation of state evolution models with the assistance of spatial information.

Inspired by the vision of integrated sensing and communications (ISAC) \cite{XiaoyangLi23, ZhaolinWang23, GuangxuZhu23}, a successive pose estimation and beam tracking (SPEBT) scheme is proposed in this paper to alleviate the beam training overhead for mmWave vehicular mobile communication systems. With the knowledge of pose information consisting of vehicle position and yaw offered by mmWave radar sensing, both BS and MS are capable of tracking the dynamically changing channel and forming directional beams to facilitate mmWave communications. To the best of our knowledge, this is the first work investigating the mmWave radar sensing-aided mmWave communications for vehicular mobile communication systems under large-scale GNSS-denied environments. The main contributions of this paper are summarized below:
\begin{enumerate}
	\item A SPEBT scheme is proposed to reduce the beam training overhead by performing mmWave beam tracking in light of pose estimation offered by mmWave radar sensing in a vehicular mobile communication system, which can normally work under extreme conditions and GNSS-denied environments.
	\item An open-source real-world mmWave sensing platform is used to verify the feasibility and effectiveness for the practical deployment of the proposed SPEBT scheme.
\end{enumerate}

\emph{Notations:} $\left(\cdot\right)\CT$, $\left(\cdot\right)\T$ and $\left(\cdot\right)^*$ stand for the Hermitian, the transpose and the conjugate operations, respectively. $|\cdot|$ represents the modulus of a complex number or the cardinality of a set. $\|\cdot\|$ represents the $\mathcal{\ell}_2$ norm of a real vector. ${\rm diag}\left(\cdot\right)$ denotes the diagonal matrix whose diagonals are the elements of the input vector. $\mathbb{SE}(2)$ and $\mathbb{SO}(2)$ stand for the 2D special Euclidean group and 2D special orthogonal group.

\section{System Model}

\subsection{MmWave Vehicular Communications Model}

\begin{figure}[t]
	\centerline{\includegraphics[width=1.0\columnwidth]{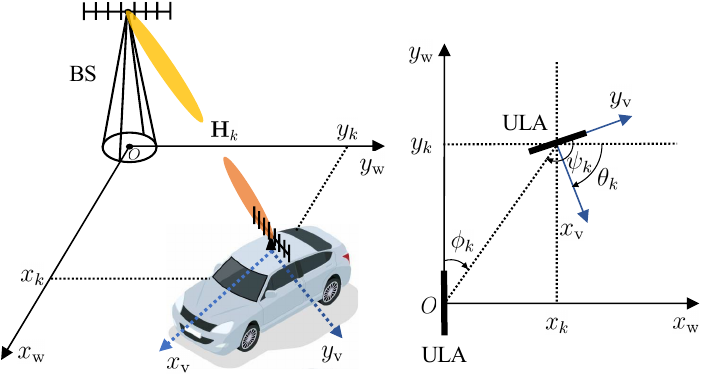}}
	\caption{mmWave vehicular communications system. \textit{Left:} 3D view of the MIMO system. \textit{Right:} 2D bird's eye view of the MIMO system.}
	\label{fig2}
\end{figure}

We consider a mmWave MIMO vehicular communications system, as illustrated in Fig. \ref{fig2}. The BS and vehicle are equipped with uniform linear arrays (ULAs) of $N$ and $M$ antennas, respectively. The mmWave wireless downlink can be described by a time-varying geometric channel model $\B{G}_k\in\mathbb{C}^{M\times N}$ as follows \cite{Akdeniz14}
\begin{align}
	\B{G}_k = \sum_{l=1}^{L_k} \alpha_{l,k} \B{a}_{\rm r}(\psi_{l,k}) \B{a}_{\rm t}\CT(\phi_{l,k})
\end{align}
where $\alpha_{l,k}$, $\phi_{l,k}$ and $\psi_{l,k}$ are the complex path gain, AoD and AoA of the $l$-th path among total $L_k$ paths at timeslot $k$. $\B{a}_{\rm t}(\cdot)\in\mathbb{C}^N$ and $\B{a}_{\rm r}(\cdot)\in\mathbb{C}^M$ are the transmit and receive array response vectors given by
\begin{align}
	\B{a}_{\rm t}(\phi) = \frac{1}{\sqrt{N}} \begin{bmatrix} 1, e^{-j \frac{2\pi d}{\lambda}\cos\phi}, \ldots, e^{-j\left(N-1\right) \frac{2\pi d}{\lambda}\cos\phi} \end{bmatrix}\T\\
	\B{a}_{\rm r}(\psi) = \frac{1}{\sqrt{M}} \begin{bmatrix} 1, e^{-j \frac{2\pi d}{\lambda}\cos\psi}, \ldots, e^{-j\left(M-1\right) \frac{2\pi d}{\lambda}\cos\psi} \end{bmatrix}\T
\end{align}
with the antenna spacing $d$ and carrier wavelength $\lambda$.

Because of the sparsity on spatial scattering in angular domain and high free-space path loss \cite{Ayach14}, mmWave channel is dominated by LoS component which is deemed as the target of beam tracking \cite{Va16, Jayaprakasam17, FuliangLiu19, Shaham20, LiChen23}. The time-varying LoS channel model $\B{H}_k\in\mathbb{C}^{M\times N}$ can be written as
\begin{align} \label{LoS Channel}
	\B{H}_k \triangleq \B{G}_k^{\rm LoS} = \alpha_k \B{a}_r(\psi_k) \B{a}_t\CT(\phi_k)
\end{align}
where the path index is omitted for simplicity. Both BS and the vehicle adopt analog beamforming to enhance directivity of transmit and receive beam patterns. The signal received by the vehicle at timeslot $k$ can be written as
\begin{align} \label{z_k}
	z_k = \B{g}_k\CT \B{H}_k \B{f}_k b_k + v_k
\end{align}
where $b_k$ and $v_k$ denote the transmitted symbol and additive
white Gaussian noise (AWGN), respectively, with $\mathbb{E}[|b_k|^2]=1$ and $v_k\sim\mathcal{CN}(0, \sigma_v^2)$. $\B{f}_k\in\mathbb{C}^N$ denotes an analog precoder at BS, and $\B{g}_k\in\mathbb{C}^M$ denotes an analog combiner at the vehicle.

\subsection{MmWave Vehicular Sensing Model}

\begin{figure}[t]
	\centerline{\includegraphics[width=1.0\columnwidth]{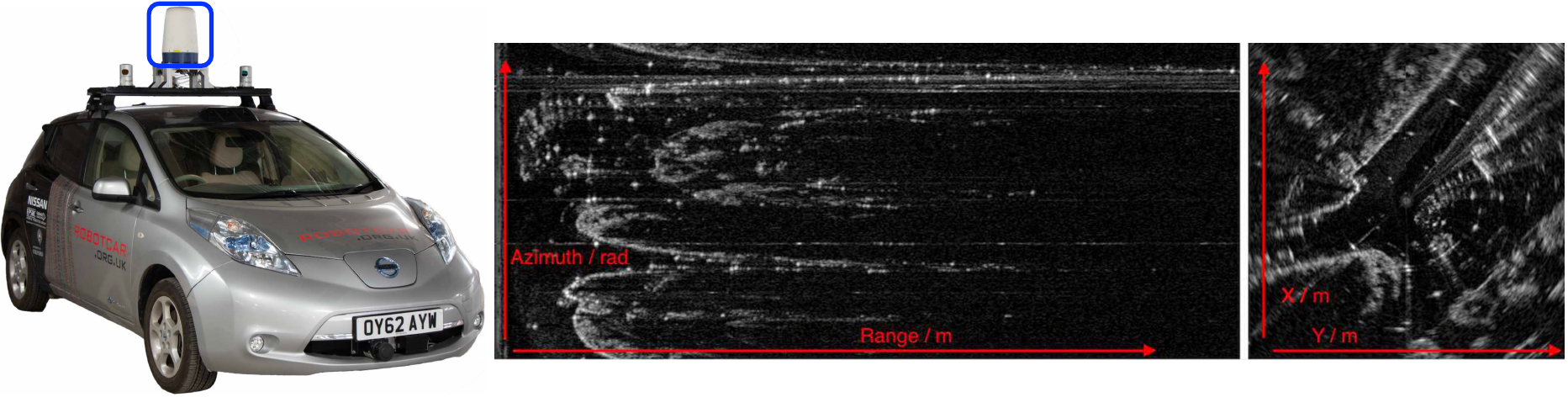}}
	\caption{mmWave vehicular sensing platform \cite{Barnes20}. \textit{Left:} a Navtech CTS350-X mmWave FMCW radar (marked with a blue box) mounted on a Nissan LEAF automobile. \textit{Middle:} a sample of raw radar point cloud in polar form. \textit{Right:} a sample of raw radar point cloud in Cartesian form.}
	\label{fig1}
\end{figure}

The mmWave vehicular mobile system applies \textit{Oxford Radar RobotCar} \cite{Barnes20} as a sensing platform to perform vehicle pose estimation with the purpose of assisting beam tracking for mmWave communications. As demonstrated in Fig. \ref{fig1}, a mmWave frequency-modulated continuous-wave (FMCW) radar sensor mounted on a vehicle emits radio waves to the ambient environment and receives radio waves reflected from the surrounding objects. Through detecting the reflective radio power returns captured by radar point cloud, the radar sensor is capable of inferring relative location (i.e., range and azimuth) of the surrounding objects with respect to the sensor itself.

As depicted in Fig. \ref{fig2}, $x_{\rm w}y_{\rm w}$-axis indicate the world coordinate frame $\underrightarrow{\mathcal{F}}_{\rm w}$, while $x_{\rm v}y_{\rm v}$-axis indicate the vehicle coordinate frame $\underrightarrow{\mathcal{F}}_{\rm v}$. By processing raw radar point cloud collected at timeslot $k$, the relative position vector $\B{r}_k = [x_k, y_k]\T$ and yaw $\theta_k$ of $\underrightarrow{\mathcal{F}}_{\rm v}$ with respect to $\underrightarrow{\mathcal{F}}_{\rm w}$ (or in other words, the absolute pose of the vehicle) can be estimated, which will be elaborated in Section \ref{S3-1}.

\section{Successive Pose Estimation and Beam Tracking}

In this section, we propose a SPEBT scheme for reducing the beam training overhead incurred in mmWave vehicular mobile communication systems. The proposed scheme involves two stages, namely 1) estimating the vehicle pose based on Fast-CFEAR approach and 2) carrying out beam tracking subsequently by employing EKF.


\subsection{Fast-CFEAR Based Pose Estimation} \label{S3-1}
Based on the state-of-the-art (SOTA) Conservative Filtering for Efficient and Accurate Radar odometry (CFEAR) framework \cite{Adolfsson23}, an approach named Fast-CFEAR is proposed to estimate the vehicle pose consisting of 2D position and yaw. The basic idea of Fast-CFEAR is to remove noise in raw radar point clouds resorting to conservative filtering, such that the underlying noise-less sparse representations can be recovered. As a result, the vehicle pose can be estimated via deploying point cloud registration between consecutive sparse representations. Fast-CFEAR is capable of offering accurate 2D position and yaw estimations with low computational complexity in real time.

The Fast-CFEAR approach can be decomposed into the following four stages:

\subsubsection{$K$-Strongest Conservative Filtering} For each azimuth of a radar point cloud in polar form, only points with the $K$ strongest pixel values exceeding a threshold $\kappa_{\rm{min}}$ are preserved. The point cloud after conservative filtering is denoted by $\mathcal{P}_{\rm F}\subset\mathbb{R}^2$. At the risk of possibly missing landmarks with low pixel values, the conservative filtering technique reduces the noise originated from false positive detections.

\subsubsection{Downsamping} Grid cells with side length $d_{\rm D}/f_{\rm D}$ are applied to downsample $\mathcal{P}_{\rm F}$ in Cartesian form. Only the centroid of each grid cell is kept in the downsampled point cloud $\mathcal{P}_{\rm D}\subset\mathbb{R}^2$. $d_{\rm D}$ is a distance parameter controlling the number of points in estimating the surface normal and $f_{\rm D}$ is a resampling factor adjusting the density of $\mathcal{P}_{\rm D}$.

\subsubsection{Estimating Oriented Surface Points} For each point $\B{p}_i = [p_i^{[1]}, p_i^{[2]}]\T\in\mathcal{P}_{\rm D}$, its all neighborhoods $\B{p}_j = [p_j^{[1]}, p_j^{[2]}]\T\in\mathcal{P}_{\rm F}$ within a radius $d_{\rm D}$ centering at $\B{p}_i$ are utilized to estimate an oriented surface point $\{\boldsymbol\mu_i, \B{n}_i\}$. The surface point $\boldsymbol\mu_i$ is computed by sample mean $\frac{1}{|\mathcal{N}_i|}\sum_{j\in\mathcal{N}_i}\B{p}_j$, where $\mathcal{N}_i = \{j\ |\ \|\B{p}_i-\B{p}_j\| \leq d_{\rm D}, \forall\B{p}_i\in\mathcal{P}_{\rm D}, \forall\B{p}_j\in\mathcal{P}_{\rm F}\}$. The surface normal $\B{n}_i$ is defined as the smallest eigenvector of sample covariance matrix
\begin{align}
	\B\Sigma_i = \frac{1}{|\B{\mathcal{N}}_i| - 1} \sum_{j\in\B{\mathcal{N}}_i}
	\begin{bmatrix}
		(p_j^{[1]}-\mu_i^{[1]})^2 & c_i\\
		c_i &(p_j^{[2]}-\mu_i^{[2]})^2
	\end{bmatrix}
\end{align}
where $c_i = (p_j^{[1]}-\mu_i^{[1]})(p_j^{[2]}-\mu_i^{[2]})$.

Through estimating the oriented surface points $\left\{\boldsymbol\mu_i, \B{n}_i\right\}$ for all points in $\mathcal{P}_{\rm D}$, a sparse representation $\mathcal{M} = \left\{\left\{\boldsymbol\mu_i, \B{n}_i\right\} |\ \forall\B{p}_i\in\mathcal{P}_{\rm D}\right\}$ of a radar point cloud can be obtained. The readers are suggested to refer to a sample sparse representation visualized by Fig. 5 in \cite{Adolfsson23} for understanding intuitively.


\subsubsection{Point Cloud Registration} Rather than resorting to multiple keyframes registration adopted in \cite{Adolfsson23} with high computational complexity, Fast-CFEAR registers the current representation $\mathcal{M}_k$ generated at $k$-th timeslot to the previous one $\mathcal{M}_{k-1}$ generated at $(k-1)$-th timeslot directly. The vehicle pose with respect to $\underrightarrow{\mathcal{F}}_{\rm w}$ at timeslot $k$ can be represented by a homogeneous transformation matrix $\B{T}_k\in\mathbb{SE}(2)$ \cite{YiMa04}
\begin{align}
	\B{T}_k = 
	\begin{bmatrix}
		\B{R}_k & \B{r}_k\\
		\B{0}\T & 1
	\end{bmatrix},\quad
	\B{R}_k = \begin{bmatrix}
		\cos\theta_k & -\sin\theta_k \\
		\sin\theta_k &  \cos\theta_k
	\end{bmatrix}
\end{align}
where $\B{R}_k\in\mathbb{SO}(2)$ is a 2D rotation matrix.

The point cloud registration can be formulated as an unconstrained minimization problem on \textit{point-to-line} (P2L) distance
\begin{align}
	&\min_{\B{T}_k}\ f_{\rm P2L} (\B{T}_k|\ \B{T}_{k-1},\mathcal{M}_{k-1}, \mathcal{M}_k)\\
	\Leftrightarrow
	&\min_{\B{r}_k, \theta_k} \sum_{\forall i\in\mathcal{M}_k} \mathcal{L}_\delta \left( \B{n}_j\T\left( \Delta\B{R}_k \boldsymbol\mu_i + \Delta\B{r}_k - \boldsymbol\mu_j \right) \right) \label{objective}
\end{align}		
which means that we seek to estimate the vehicle pose $\{\hat{\B{r}}_k, \hat{\theta}_k\}$ at timeslot $k$ for best registering the consecutive sparse representations $\mathcal{M}_{k-1}$ and $\mathcal{M}_k$ given the vehicle pose $\{\B{r}_{k-1}, \theta_{k-1}\}$ at timeslot $k-1$.

The objective function in (\ref{objective}) is formulated by taking a summation over P2L distances between the surface points $\boldsymbol\mu_i\in\mathcal{M}_k$ and their neighborhoods $\boldsymbol\mu_j\in\mathcal{M}_{k-1}$, where Huber loss $\mathcal{L}_\delta(\cdot)$ is used to reduce the objective's sensitivity against outliers. The relative vehicle pose from timeslot $k-1$ to timeslot $k$ can be computed by
\begin{align}
	\B{T}_{k-1}^{-1} \B{T}_k = 
	\begin{bmatrix}
		\Delta\B{R}_k & \Delta\B{r}_k\\
		\B{0}\T & 1
	\end{bmatrix},
	\Delta\B{R}_k = \begin{bmatrix}
		\cos\Delta\theta_k & -\sin\Delta\theta_k \\
		\sin\Delta\theta_k & \cos\Delta\theta_k
	\end{bmatrix}
\end{align}
where $\Delta\B{r}_k = \B{r}_k - \B{r}_{k-1}$ is the relative position vector and $\Delta\B{R}_k\in\mathbb{SO}(2)$ is the relative rotation matrix with relative yaw $\Delta\theta_k=\theta_k-\theta_{k-1}$.

For each surface point $\boldsymbol\mu_i\in\mathcal{M}_k$, its neighborhoods $\boldsymbol\mu_j\in\mathcal{M}_{k-1}$ must satisfies two conditions: firstly, the angle between surface normals $\B{n}_i$ and $\B{n}_j$ is within a tolerance $\Theta$; secondly, $\boldsymbol\mu_j$ should be the closest point to $\boldsymbol\mu_i$ within the radius $d_{\rm D}$ centering at $\boldsymbol\mu_i$.

The proposed Fast-CFEAR approach applies a non-linear optimization library Ceres Solver \cite{CeresSolver} to numerically solve the unconstrained minimization problem (\ref{objective}) using BFGS line search method.

\subsection{EKF Based Beam Tracking}

\begin{figure*}[t]
	\begin{align} 
		\B{w}_k = \begin{bmatrix}
			\frac{\gamma}{2}\hat{\rho}_k|\alpha_{k-1}|\left( \frac{\hat{\B{r}}_k\T\BS{\epsilon}_k^r}{||\hat{\B{r}}_k||^2} - \frac{\hat{\B{r}}_{k-1}\T\BS{\epsilon}_{k-1}^r}{||\hat{\B{r}}_{k-1}||^2} \right),&
			\frac{2\pi}{\lambda} \left( \frac{\hat{\B{r}}_{k-1}\T\BS{\epsilon}_{k-1}^r}{||\hat{\B{r}}_{k-1}||} - \frac{\hat{\B{r}}_k\T\BS{\epsilon}_k^r}{||\hat{\B{r}}_k||} \right),&
			\frac{\tilde{\B{r}}_{k-1}\T\BS{\epsilon}_{k-1}^r}{||\hat{\B{r}}_{k-1}||^2} - \frac{\tilde{\B{r}}_k\T\BS{\epsilon}_k^r}{||\hat{\B{r}}_k||^2},&
			\frac{\tilde{\B{r}}_{k-1}\T\BS{\epsilon}_{k-1}^r}{||\hat{\B{r}}_{k-1}||^2} - \frac{\tilde{\B{r}}_k\T\BS{\epsilon}_k^r}{||\hat{\B{r}}_k||^2} + \Delta\B{\epsilon}_k^\theta
		\end{bmatrix}\T
		\label{w_k}
	\end{align}
	\hrulefill
\end{figure*}

The EKF based beam tracking algorithm is composed of state evolution model and measurement model, which correspond to channel prediction and update stages, respectively.

\subsubsection{State Evolution Model} The channel state vector can be defined as
\begin{align}
	\B{s}_k = 
	\begin{bmatrix}
		|\alpha_k|,\ \arg\alpha_k,\ \phi_k,\ \psi_k
	\end{bmatrix}\T
\end{align}
where $|\alpha_k|$ and $\arg\alpha_k$ denote the magnitude and argument of the path gain. The channel state evolution model can be represented by
\begin{align} \label{state_evolution}
	\B{s}_k = \B{f}(\B{s}_{k-1}) + \B{u}_k + \B{w}_k
\end{align}
where $\B{f}(\cdot)$ is the evolution function. $\B{u}_k$ and $\B{w}_k$ denote the control input and process noise, respectively.

\textit{Proposition 1 (Ground Truth Pose Based Channel State Evolution Model):} Based on ground truth of the vehicle pose, the channel state evolution model can be established as
\begin{align}
	\B{s}_k = \B{F}_k \B{s}_{k-1} + \B{u}_k
\end{align}
where $\B{F}_k = {\rm diag}([\rho_k,1,1,1])$ is an evolution matrix with $\rho_k = (||\B{r}_{k-1}||/||\B{r}_k||)^{\frac{\gamma}{2}}$ standing for correlation coefficient on the magnitude of path gain. The path loss exponent is given by $\gamma$. $\B{u}_k = [0, \frac{2\pi}{\lambda}(||\B{r}_k||-||\B{r}_{k-1}||), u_k^{[3]}, u_k^{[3]}- \Delta\theta_k]\T$ is the deterministic control input, where $u_k^{[3]} = \arctan(\frac{x_k}{y_k}) - \arctan(\frac{x_{k-1}}{y_{k-1}})$.

\emph{Proof:} See Appendix \ref{proof_1}. $\hfill\blacksquare$

Nevertheless, the ground truth of vehicle pose is not available in practice. Instead, the vehicle pose estimation is accessible as we elaborated in Section \ref{S3-1}. Taking pose estimation error into consideration, another channel state evolution model can be derived by the following proposition.

\textit{Proposition 2 (Pose Estimation Based Channel State Evolution Model):} Based on the vehicle pose estimation, the channel state evolution model can be established as
\begin{align}
	\B{s}_k \approx \hat{\B{F}}_k \B{s}_{k-1} + \hat{\B{u}}_k + \B{w}_k
\end{align}
where $\hat{\B{F}}_k$ and $\hat{\B{u}}_k$ are estimated evolution matrix and control input, which have the same expressions as $\B{F}_k$ and $\B{u}_k$ but are calculated with the pose estimation instead of ground truth. $\B{w}_k\sim\mathcal{N}(\B{0}, \B{Q}_k)$ is the process noise given in (\ref{w_k}) which is placed on the top of this page. The covariance $\B{Q}_k$ of the process noise is given by
\begin{align}
	&\B{Q}_k = \begin{bmatrix}
		\B{A}_k&\B{0}\\
		\B{0}&\B{B}_k
	\end{bmatrix},\quad
	\B{B}_k = \beta_k\sigma_r^2
	\begin{bmatrix}
		1&1\\
		1&1 + \frac{2\sigma_\theta^2}{\beta_k\sigma_r^2}
	\end{bmatrix}\\
	&\B{A}_k = \sigma_r^2
	\begin{bmatrix}
		\left(\frac{\gamma}{2}\hat{\rho}_k|\alpha_{k-1}|\right)^2\beta_k&
		-\frac{\pi}{\lambda}\gamma\hat{\rho}_k|\alpha_{k-1}|\zeta_k\\
		-\frac{\pi}{\lambda}\gamma\hat{\rho}_k|\alpha_{k-1}|\zeta_k&\left(\frac{2\pi}{\lambda}\right)^2
	\end{bmatrix}
\end{align}
where $\beta_k = \frac{1}{||\hat{\B{r}}_{k-1}||^2} + \frac{1}{||\hat{\B{r}}_k||^2}$ and $\zeta_k = \frac{1}{||\hat{\B{r}}_{k-1}||} + \frac{1}{||\hat{\B{r}}_k||}$.

\emph{Proof:} See Appendix \ref{proof_2}. $\hfill\blacksquare$

\textit{Remark 1:} For mmWave channel, the argument of path gain $\arg\alpha_k$ cannot be accurately tracked even based on mm-level localization techniques.

\emph{Proof:} The standard deviation on tracking $\arg\alpha_k$ is given by $(\B{Q}_k^{[2,2]})^{\frac{1}{2}} = \frac{2\pi}{\lambda}\sigma_r$ in Proposition 2, which ranges approximately $600\sigma_r \sim 6000\sigma_r\ {\rm rads}$ for mmWave channel. Even with localization techniques of mm-level accuracy (i.e., $\sigma_r\approx10^{-3}{\rm m}$), the standard deviation on tracking $\arg\alpha_k$ still ranges about $0.6\sim6.0\ {\rm rads}$ (i.e., $34^\circ\sim340^\circ$). Such a large standard deviation indicates that the estimation on $\arg\alpha_k$ is always inaccurate. $\hfill\blacksquare$

Leaving the tracking on $\arg\alpha_k$ out of account, the channel state vector can be re-defined as
\begin{align}
	\bar{\B{s}}_k = 
	\begin{bmatrix}
		|\alpha_k|,\ \phi_k,\ \psi_k
	\end{bmatrix}\T
\end{align}
and the pose estimation based channel state evolution model can be modified accordingly
\begin{align}
	\bar{\B{s}}_k \approx \bar{\B{F}}_k \bar{\B{s}}_{k-1} + \bar{\B{u}}_k + \bar{\B{w}}_k
\end{align}
where $\bar{\B{F}}_k$, $\bar{\B{u}}_k$ and $\bar{\B{w}}_k$ are calculated by $\hat{\B{F}}_k$, $\hat{\B{u}}_k$ and $\B{w}_k$ excluding the $2^{\rm nd}$ row (and the $2^{\rm nd}$ column) relevant to $\arg\alpha_k$.

\begin{algorithm}[t]
	\caption{The Proposed SPEBT Scheme}
	\begin{algorithmic}[1]
		\State \textbf{input:} $\B{r}_0$, $\theta_0$, $\mathcal{M}_0$
		\State \textbf{output:} $\bar{\B{s}}_{k|k} = 
		\begin{bmatrix}
			|\alpha_k|,\ \phi_k,\ \psi_k
		\end{bmatrix}\T$
		\State \textbf{initialization:} Set timeslot $k=0$. Initialize channel state vector $\bar{\B{s}}_{\rm I}$ via mmWave initial access techniques. Set $\bar{\B{s}}_{0|0}$ = $\bar{\B{s}}_{\rm I}$ and $\B{P}_{0|0}=\bar{\B{Q}}_0$. Set thresholds $\tilde{\alpha}$, $\tilde{\phi}$, $\tilde{\psi}>0$.
		\Repeat $\ k\leftarrow k+1$
		\Statex \quad\enspace \textit{// \textbf{Fast-CFEAR Based Pose Estimation}}
		\State The latest radar point cloud is available at timeslot $k$.
		\State Obtain a sparse representation $\mathcal{M}_k$ by performing the first three stages of Fast-CFEAR approach.
		\State Estimate the vehicle pose $(\hat{\B{r}}_k, \hat{\theta}_k)$ via point cloud registration.
		\Statex \quad\enspace \textit{// \textbf{EKF Based Beam Tracking}}
		\State Beam tracking via configuring the analog precoder $\B{f}_k = \B{a}_{\rm t}(\phi_{k-1})$ and the analog combiner $\B{g}_k = \B{a}_{\rm r}(\psi_{k-1})$.
		\Statex \quad\enspace -- \textit{Channel Prediction Stage:}
		\State A Priori Channel State Vector:
		\Statex \quad\enspace $\bar{\B{s}}_{k|k-1} = \bar{\B{F}}_k\bar{\B{s}}_{k-1|k-1} + \bar{\B{u}}_k$
		\State A Priori Channel Covariance Matrix:
		\Statex \quad\enspace $\B{P}_{k|k-1} = \bar{\B{F}}_k\B{P}_{k-1|k-1}\bar{\B{F}}_k\T + \bar{\B{Q}}_k$
		\Statex \quad\enspace -- \textit{Channel Update Stage:}
		\State Kalman Gain:
		\Statex \quad\enspace $\B{K}_k = \frac{\B{P}_{k|k-1}\nabla\bar{h}(\bar{\B{s}}_{k|k-1})}{\nabla\T\bar{h}(\bar{\B{s}}_{k|k-1}) \B{P}_{k|k-1} \nabla\bar{h}(\bar{\B{s}}_{k|k-1}) + \sigma_{\tilde{v}_k}^2}$
		\State A Posteriori Channel State Vector:
		\Statex \quad\enspace $\bar{\B{s}}_{k|k} = \bar{\B{s}}_{k|k-1} + \B{K}_k(\bar{z}_k - \nabla\T\bar{h}(\bar{\B{s}}_{k|k-1})\bar{\B{s}}_{k|k-1})$
		\State A Posteriori Channel Covariance Matrix:
		\Statex \quad\enspace $\B{P}_{k|k} = (\B{I} - \B{K}_k\nabla\T\bar{h}(\bar{\B{s}}_{k|k-1}))\B{P}_{k|k-1}$
		\Statex \quad\enspace -- \textit{Deviation Checking for Re-initialization:}
		\State \textbf{if} $||\alpha_k| - |\alpha_{\rm I}|| > \tilde{\alpha}$ or $|\phi_k - \phi_{\rm I}| > \tilde{\phi}$ or $|\psi_k - \psi_{\rm I}| > \tilde{\psi}$
		\State \qquad Re-initialize $\bar{\B{s}}_{\rm I}$ and set $\bar{\B{s}}_{k|k}$ = $\bar{\B{s}}_{\rm I}$.
		\State \textbf{end}
		\Until The vehicle arrives at destination.
	\end{algorithmic}
	\label{alg}
\end{algorithm}

\subsubsection{Measurement Model} At each timeslot $k$, a pilot symbol $b_k$ undergoing the downlink channel $\B{H}_k$ is transmitted from BS to the vehicle. As given in (\ref{z_k}), a complex-valued measurement model can be established as
\begin{align}
	z_k = h(\B{s}_k) + v_k
\end{align}
where $h(\B{s}_k) = \alpha_k\B{g}_k\CT\B{a}_r(\psi_k)\B{a}_t\CT(\phi_k)\B{f}_kb_k$ is the complex-valued measurement function. Nonetheless, a real-valued measurement model is still in need for the convenience of deploying EKF algorithm that generally works in real field. Since the phase information of measurement $z_k$ is inessential as the tracking on $\arg\alpha_k$ is not involved, a real-valued measurement model can be obtained by taking the magnitude of $z_k$ as follows
\begin{align}\label{z_bar_k}
	\bar{z}_k = \bar{h}(\bar{\B{s}}_k) + \tilde{v}_k
	\approx \bar{h}(\bar{\B{s}}_k) + \bar{v}_k
\end{align}
where the real-valued measurement and measurement function are given by $\bar{z}_k = |z_k|^2 - \sigma_v^2$ and $\bar{h}(\bar{\B{s}}_k) \triangleq |h(\B{s}_k)|^2 = |b_k|^2 |\alpha_k|^2 |\B{g}_k\CT\B{a}_r(\psi_k)|^2 |\B{f}_k\CT\B{a}_t(\phi_k)|^2$, respectively. The measurement noise is given by $\tilde{v}_k = 2\Re[h(\B{s}_k) v_k^*] + |v_k|^2 - \sigma_v^2$ with zero mean and variance $\sigma_{\tilde{v}_k}^2 = (\sigma_v^2 + 2\bar{h}(\bar{\B{s}}_k))\sigma_v^2$.
 
Last but not the least, it can be observed that $\tilde{v}_k$ is no longer a Gaussian random variable. Instead, $\tilde{v}_k$ is a summation of a Gaussian random variable and a chi-squared random variable with $2$ degrees of freedom. For the sake of guaranteeing the assumption of Gaussian measurement noise in EKF algorithm, a Gaussian noise $\bar{v}_k\sim\mathcal{N}(0, \sigma_{\tilde{v}_k}^2)$ with the same variance as $\tilde{v}_k$ is adopted to substitute the non-Gaussian $\tilde{v}_k$ in the final measurement model (\ref{z_bar_k}).

\begin{table}[t]
	\centering
	\caption{Simulation Parameters}
	\resizebox{1.0\columnwidth}{!}{%
		\begin{tabular}{|l|l|}
			\hline
			\textbf{Pose Estimation Subsystem} & \textbf{Parameter Value}\\
			\hline
			Selected sensing range & $5{\rm m}\sim100{\rm m}$\\
			\hline
			$K$-strongest filtering & $K=3$\\
			\hline
			Pixel value threshold (integer: $0\sim255$)& $\kappa_{\rm min}=55$\\
			\hline
			Side length of downsampling grid cell & $d_{\rm D}=3.5{\rm m}$\\
			\hline
			Resampling factor & $f_{\rm D}=1.0$\\
			\hline
			Angle tolerance & $\Theta=30^\circ$\\
			\hline
			Huber loss $\mathcal{L}_\delta$ & $\delta=0.1$\\
			\hline
			\hline
			\textbf{Beam Tracking Subsystem} & \textbf{Parameter Value}\\
			\hline
			Carrier frequency and wavelength & $f=50{\rm GHz}$, $\lambda=6{\rm mm}$\\
			\hline
			Number of transmit/receive antennas & $N=M=4$\\
			\hline
			Antenna spacing & $d=\lambda/2$\\ 
			\hline
			Path loss exponent & $\gamma=2.2$\\
			\hline
			Magnitude of path gain at $d_0=1{\rm m}$ & $|\alpha_{\rm ref}| = 5\times10^{-4}$\\
			\hline
			Noise power & $\sigma_v^2=-90{\rm dBm}$\\
			\hline
			Standard deviation of estimation error & $\sigma_r=1.0{\rm m}, \sigma_\theta=3^\circ$\\
			\hline
			Channel re-initialization thresholds &  \makecell[l]{$\tilde{\alpha}=5\times 10^{-7}$\\ $\tilde{\phi}=\tilde{\psi}=7.5^\circ$}\\
			\hline
			Coordinate of vehicle initial position & $(0{\rm m},0{\rm m})$\\
			\hline
			Coordinate of BS position & $(-30{\rm m},-125{\rm m})$\\
			\hline
		\end{tabular}
	}
	\label{parameter}
\end{table}

\subsection{The Proposed SPEBT Scheme}

The SPEBT scheme is summarized in Algorithm \ref{alg}. At the start of the scheme, the initial channel state vector $\bar{\B{s}}_{\rm I}$ should be established via mmWave initial access techniques \cite{Giordani16}, which is out of the scope of this paper. The vehicle performs beam tracking in light of the pose estimation whenever the latest point cloud collected by the radar sensor is available. To guarantee the accuracy and timeliness of the tracked channel, the algorithm should re-initialize the channel state vector once the deviation between the tracked channel and the initial channel are greater than the given thresholds. At the channel update stage, $\nabla\bar{h}(\cdot)$ represents the gradient of the measurement function with respect to the channel state vector, whose expression is provided in Appendix \ref{gradient}.

\section{Simulation Results}

\begin{figure}[t]
	\centering
	\includegraphics[width=1.0\columnwidth]{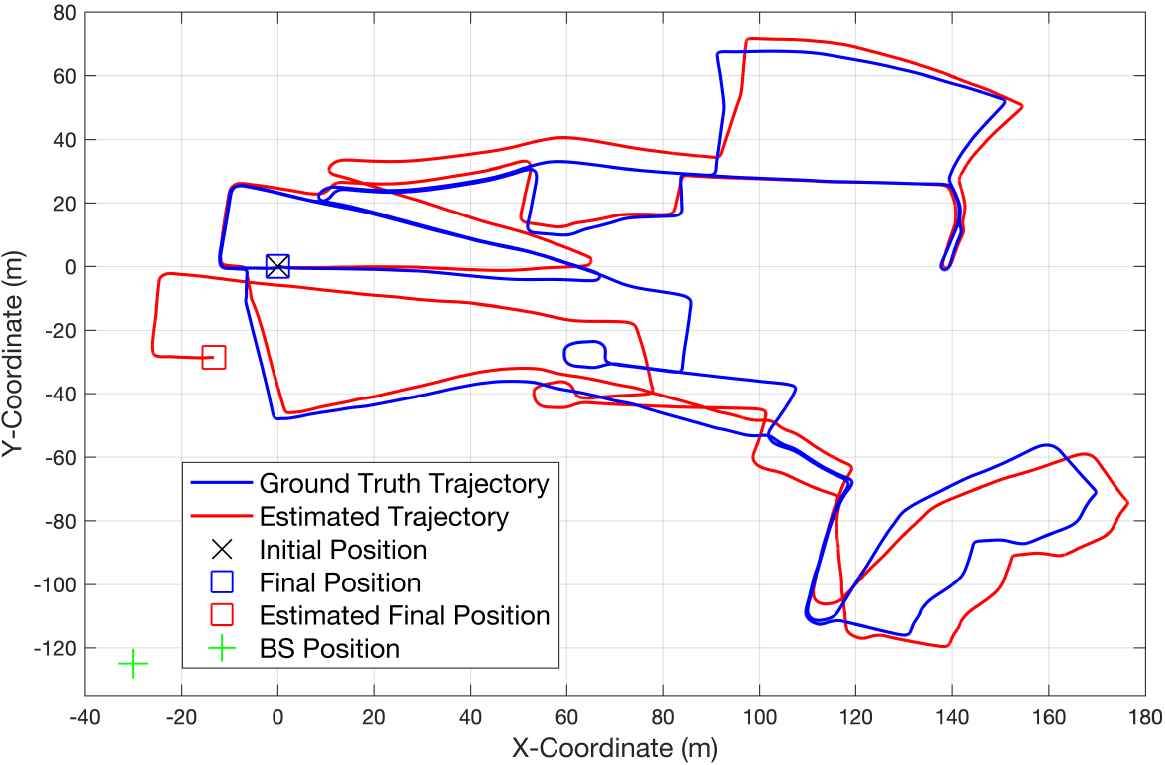}
	\caption{The estimated vehicle moving trajectory compared to ground truth. Both trajectories are aligned by the initial position marked with $\times$, and the final positions are marked with $\square$. The position of BS is marked with $+$.}
	\label{trajectory}
\end{figure}

In this section, experimental simulations are carried out to validate the performance of the proposed SPEBT scheme. For the simplicity of implementation, the simulation follows a divide and conquer strategy using two subsystems: a pose estimation subsystem and a pose estimation based beam tracking subsystem. On the one hand, as depicted in Fig. \ref{fig1}, a sequence named \textit{2019-01-15-14-24-38} in \textit{Oxford Radar RobotCar} dataset is selected to perform vehicle pose estimation. The sequence is consisted of 6388 consecutive frames collected by a Navtech CTS350-X mmWave FMCW radar in an urban environment. This subsystem is implemented by C++ programming language in Visual Studio 2022 for the sake of real-time processing. On the other hand, as illustrated by the 2D bird's eye view in Fig. \ref{fig2}, the beam tracking is performed in a 2D mmWave MIMO mobile wireless system, and this subsystem is implemented in MATLAB R2023a regarding the pose estimation information as input. The simulation parameters of both subsystems are summarized in Table \ref{parameter}.

\begin{figure}[t]
	\centering
	\subfigure{
		\begin{minipage}[t]{1\linewidth}
			\centering
			\includegraphics[width=0.95\columnwidth]{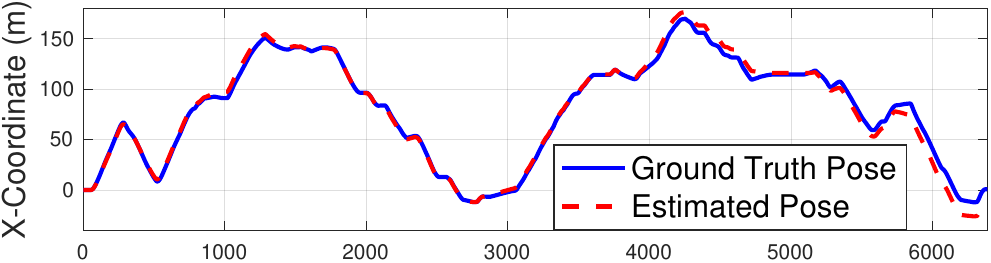}
		\end{minipage}}
	\subfigure{
		\begin{minipage}[t]{1\linewidth}
			\centering
			\includegraphics[width=0.95\columnwidth]{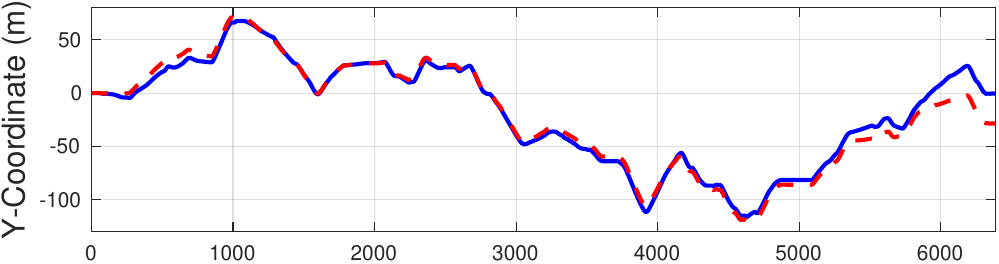}
		\end{minipage}}
	\subfigure{
		\begin{minipage}[t]{1\linewidth}
			\centering
			\includegraphics[width=0.95\columnwidth]{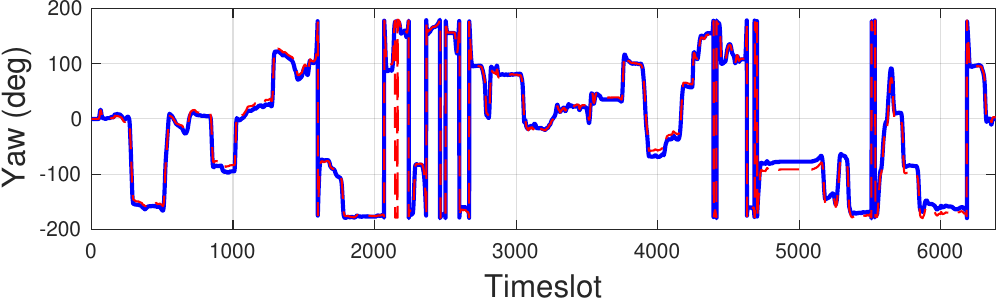}
		\end{minipage}}
	\centering
	\caption{The estimated pose compared to ground truth pose on coordinate $(x_k,y_k)$ and yaw $\theta_k$ against timeslot $k$.}
	\label{pose_estimation}
\end{figure}

In Fig. \ref{trajectory}, we demonstrate the estimated vehicle moving trajectory and ground truth trajectory, respectively. In addition, Fig. \ref{pose_estimation} compares the estimated pose with ground truth pose on the coordinate and yaw against the timeslot. It can be observed from both figures that the estimations are in high accuracy at the start of the trajectory, and they gradually deviate from the ground truth as the vehicle moving due to the accumulation of estimation errors. As a consequence, an obvious disparity is produced at vehicle final position. This observation brings an insight that it is crucial to take the pose estimation error into account in the beam tracking modeling as we elaborated in Proposition 2. That is also the reason why we cannot apply the simpler model offered by Proposition 1 directly. Table \ref{pose_error} provides the vehicle pose estimation evaluated by root-mean-squared error (RMSE) metric and KITTI benchmark \cite{KITTI} respectively, which are generated by an open-source toolbox \cite{Toolbox} that is commonly adopted in pose evaluation. The statistics confirm that our pose estimation results are comparable to those of the SOTA approaches.

\begin{figure}[t]
	\centering
	\subfigure{
		\begin{minipage}[t]{1\linewidth}
			\centering
			\includegraphics[width=0.95\columnwidth]{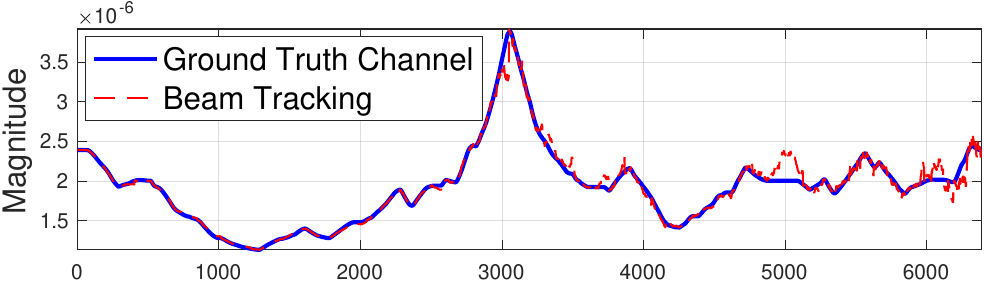}
	\end{minipage}}
	\subfigure{
		\begin{minipage}[t]{1\linewidth}
			\centering
			\includegraphics[width=0.95\columnwidth]{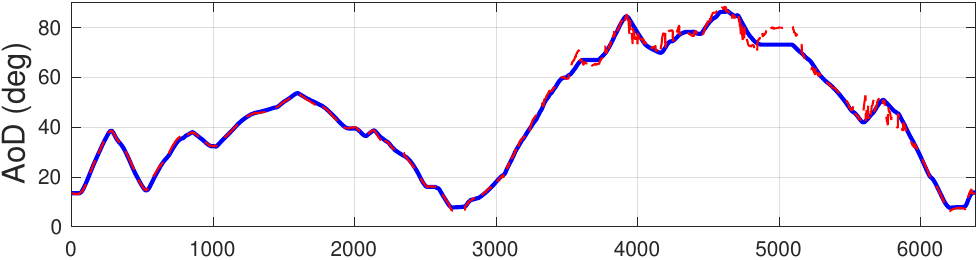}
	\end{minipage}}
	\subfigure{
		\begin{minipage}[t]{1\linewidth}
			\centering
			\includegraphics[width=0.95\columnwidth]{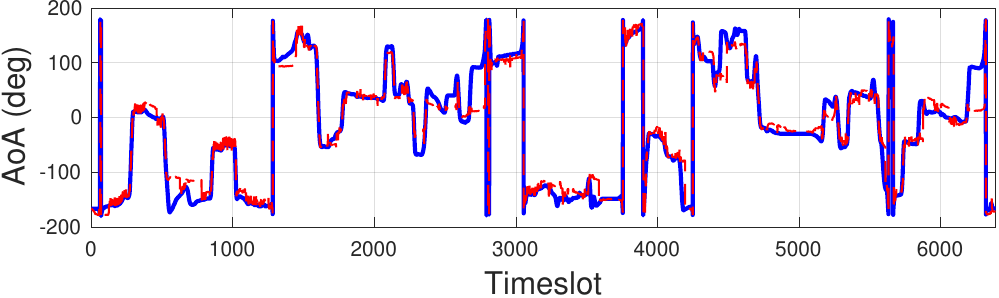}
	\end{minipage}}
	\centering
	\caption{A beam tracking realization generated by SPEBT scheme compared to ground truth channel on $|\alpha_k|$, $\phi_k$ and $\psi_k$ against timeslot $k$.}
	\label{beam_tracking}
\end{figure}

In Fig. \ref{beam_tracking}, we compare a beam tracking realization generated by SPEBT scheme with ground truth channel on the magnitude of path gain, AoD and AoA against the timeslots, respectively. Similarly, the tracking on magnitude is almost in accordance with ground truth in the first half 3000 timeslots, but it fluctuates intensively around ground truth in the second half duration, which coincides with the gradual reduction in confidence of pose estimations. The beam tracking performance evaluated by RMSE is provided in Table \ref{tracking_error}, which is the average result of $10^5$ i.i.d. Monte Carlo simulations. From both the qualitative and quantitative results in Fig. \ref{beam_tracking} and Table \ref{tracking_error}, it is interesting to find that the tracking on AoD performs much better than the tracking on AoA. It is because that AoD is only related to the vehicle position owing to the stationary BS antenna. However, AoA is additionally relevant to the vehicle yaw since the orientation of vehicular antenna keeps changing.

Last but not the least, it is satisfying to point out that the average times on channel re-initialization is less than 300 out of the total 6388 timeslots, which means that the proposed SPEBT scheme can greatly reduce the proportion of beam training overhead to less than $5\%$ for mmWave vehicular mobile communication systems.

\begin{table}[t]
	\centering
	\caption{Vehicle Pose Estimation Error}
	\resizebox{1.0\columnwidth}{!}{%
		\begin{tabular}{|l|l|l|l|}
			\hline
			\multicolumn{2}{|l|}{\textbf{Position Estimation Error}} & \multicolumn{2}{l|}{\textbf{Yaw Estimation Error}}\\
			\hline
			RMSE ($\rm m$) & KITTI ($\%$) & RMSE ($\deg$) & KITTI ($\deg/{\rm m}$)\\
			\hline
			$(2.772, 4.083)$ & $6.562$ & $3.550$ & $0.0125$\\
			\hline
		\end{tabular}
	}
	\label{pose_error}
\end{table}

\begin{table}[t]
	\centering
	\caption{Beam Tracking RMSE}
	\resizebox{1.0\columnwidth}{!}{%
		\begin{tabular}{|l|l|l|l|}
			\hline
			\textbf{Channel} & Path Gain Mag. ($\%$)& AoD ($\deg$) & AoA ($\deg$)\\
			\hline
			\textbf{RMSE} & 1.580 & 0.220 & 4.616\\
			\hline
		\end{tabular}
	}
	\label{tracking_error}
\end{table}

\section{Conclusion}

We studied the mmWave radar sensing-aided mmWave communications aiming at reducing beam training overhead in high mobility vehicular mobile communication systems. A SPEBT scheme is proposed to achieve this goal, and simulations on an open-source real-world sensing dataset are conducted to verify its feasibility and effectiveness. A potential future direction is to extend the proposed scheme on 3-dimensional (3D) sensing dataset so as to generate precise 3D pose estimation results and accurate 3D beam tracking output.

\appendices

\section{Proof of Proposition 1}\label{proof_1}
Based on simplified path loss model \cite{Goldsmith05}, the complex path gain of LoS channel is formulated as
\begin{align}
	\alpha = \alpha_{\rm ref} \left(d_0/||\B{r}||\right)^\frac{\gamma}{2} e^{j\frac{2\pi}{\lambda}\left(||\B{r}|| - d_0\right)}
\end{align}
where $\alpha_{\rm ref}$ is the complex path gain at reference distance $d_0$. Therefore, the magnitude and argument of path gain are $|\alpha| = |\alpha_{\rm ref}|(d_0/||\B{r}||)^\frac{\gamma}{2}$ and $\arg\alpha = \arg\alpha_{\rm ref} + \frac{2\pi}{\lambda}(||\B{r}||-d_0)$, respectively, which lead to the evolution equations $|\alpha_k| = \rho_k|\alpha_{k-1}|$ and $\arg\alpha_k = \arg\alpha_{k-1} + \frac{2\pi}{\lambda}(||\B{r}_k||-||\B{r}_{k-1}||)$. The derivation on the evolution equation of AoD is straightforward based on the 2D geometry illustrated in Fig. \ref{fig2}. Besides, the relation between yaw, AoD and AoA is described by $\theta-\phi+\psi=\pi$, which leads to the evolution equation of AoA accordingly.

\section{Proof of Proposition 2}\label{proof_2}
The pose estimations are generally modeled as
\begin{align}
	\hat{\B{r}}_k = \B{r}_k + \BS{\epsilon}_k^r,\quad
	\hat\theta_k = \theta_k + \epsilon_k^\theta
\end{align}
where $\hat{\B{r}}_k$ and $\hat\theta_k$ are estimations on the position vector $\B{r}_k$ and yaw $\theta_k$, and $\BS{\epsilon}_k^r \sim \mathcal{N}(\B{0}, \sigma_r^2 \B{I}_2)$ and $\epsilon_k^\theta \sim \mathcal{N}(0, \sigma_\theta^2)$ are the corresponding estimation errors. Since the distance $||\B{r}_k||$ between the two transceivers and the absolute value of yaw $|\theta_k|$ are usually much larger than the estimation errors $||\BS{\epsilon}_k^r||$ and $|\epsilon_k^\theta|$, respectively, the evolution equation of AoA can be approximated by the first-order Taylor series as follows
\begin{align}
	\psi_k
	&= f^{[4]}(\B{r}_{k-1}, \B{r}_k, \theta_{k-1}, \theta_k, \psi_{k-1})\\
	&\approx f^{[4]}(\hat{\B{r}}_{k-1}, \hat{\B{r}}_k, \hat{\theta}_{k-1}, \hat{\theta}_k, \psi_{k-1}) + w_k^{[4]}
\end{align}
where $f^{[4]}(\cdot)$ is the evolution equation of AoA given in Proposition 1. The process noise $w_k^{[4]} = -(\frac{\partial f^{[4]}}{\partial\B{r}_{k-1}}|_{\hat{\B{r}}_{k-1}})\T\BS{\epsilon}_{k-1}^r - (\frac{\partial f^{[4]}}{\partial\B{r}_k}|_{\hat{\B{r}}_k})\T\BS{\epsilon}_k^r - (\frac{\partial f^{[4]}}{\partial\theta_{k-1}}|_{\hat\theta_{k-1}})\T\BS{\epsilon}_{k-1}^\theta - (\frac{\partial f^{[4]}}{\partial\theta_k}|_{\hat\theta_k})\T\BS{\epsilon}_k^\theta$ is consisted of the first-order Taylor polynomials and its final expression is provided by the fourth entry of (\ref{w_k}), where $\tilde{\B{r}}_k = [\hat{y}_k, -\hat{x}_k]\T$ and $\Delta\epsilon_k^\theta = \epsilon_k^\theta - \epsilon_{k-1}^\theta$. In addition, $w_k^{[4]}$ is a zero mean Gaussian random variable with covariance $\B{Q}_k^{[4,4]}$ (i.e., $w_k^{[4]}\sim\mathcal{N}(0, \beta_k\sigma_r^2 + 2\sigma_\theta^2)$) since it is a linear combination of zero mean Gaussian random variables. Besides, the first-order Taylor approximations on the evolution equations of other channel states can be derived in the same way, which are omitted here for avoidance of duplication.

\section{Gradient of Measurement Function}\label{gradient}
The gradient $\nabla\bar{h}(\cdot)$ of the measurement function $\bar{h}(\cdot)$ with respect to the channel state vector $\bar{\B{s}}_k$ can be written as
\begin{align}
	\nabla\bar{h}(\bar{\B{s}}_k) =
	\begin{bmatrix}
		2|b_k|^2|\alpha_k||\B{g}_k\CT\B{a}_r(\psi_k)|^2|\B{f}_k\CT\B{a}_t(\phi_k)|^2\\
		\frac{2\pi d\sin\phi_k}{N\lambda}|b_k|^2|\alpha_k|^2|\B{g}_k\CT\B{a}_r(\psi_k)|^2|\B{f}_k\CT \B{\Phi}_k\B{f}_k\\
		\frac{2\pi d\sin\psi_k}{M\lambda}|b_k|^2|\alpha_k|^2|\B{f}_k\CT\B{a}_t(\phi_k)|^2\B{g}_k\CT \B{\Psi}_k\B{g}_k
	\end{bmatrix}
\end{align}
where $\B{\Phi}_k\in\mathbb{C}^{N\times N}$ has the following expression
\begin{align}
	\B{\Phi}_k =
	\begin{bmatrix}
		\eta_0&\eta_1^*&\eta_2^*&\ldots&\eta_{N-2}^*&\eta_{N-1}^*\\
		\eta_1&\eta_0&\eta_1^*&\ldots&\eta_{N-3}^*&\eta_{N-2}^*\\
		\eta_2&\eta_1&\eta_0&\ldots&\eta_{N-4}^*&\eta_{N-3}^*\\
		\vdots&\vdots&\vdots&\ddots&\vdots&\vdots\\
		\eta_{N-2}&\eta_{N-3}&\eta_{N-4}&\ldots&\eta_0&\eta_1^*\\
		\eta_{N-1}&\eta_{N-2}&\eta_{N-3}&\ldots&\eta_1&\eta_0
	\end{bmatrix}
\end{align}
with each entry specified by $\eta_n =ne^{-j(n\frac{2\pi d}{\lambda}\cos\phi_k-\frac{\pi}{2})}, n = 0,1,\ldots,N-1$. $\B{\Psi}_k\in\mathbb{C}^{M\times M}$ is of the same form with $\eta_n$ replaced by $\xi_m = me^{-j(m\frac{2\pi d}{\lambda}\cos\psi_k-\frac{\pi}{2})}, m = 0,1,\ldots,M-1$.

\ifCLASSOPTIONcaptionsoff
\newpage
\fi

\bibliographystyle{IEEEtran}

\bibliography{Ref_BeamTracking}

\end{document}